\documentclass[twocolumn,showpacs,preprintnumbers,amsmath,amssymb, prb, reprint]{revtex4-1}

\usepackage{stmaryrd}
\usepackage{txfonts}
\usepackage{amssymb}
\usepackage{mathrsfs}
\usepackage{graphicx}
\usepackage{dcolumn}
\usepackage{bm}
\usepackage{epsfig}
\usepackage{color}                    
\usepackage{hyperref}                 

\begin{document}
\preprint{\href{http://dx.doi.org/10.1103/PhysRevB.87.214419}{S.-Z. Lin, C. Reichhardt, C. D. Batista and A. Saxena , Phys. Rev. B {\bf 87}, 214419 (2013).}}

\title{Particle model for skyrmions in metallic chiral magnets: dynamics, pinning and creep}
\author{Shi-Zeng Lin}
\affiliation{Theoretical Division, Los Alamos National Laboratory, Los Alamos, New Mexico 87545, USA}

\author{ Charles Reichhardt}
\affiliation{Theoretical Division, Los Alamos National Laboratory, Los Alamos, New Mexico 87545, USA}

\author{Cristian D. Batista}
\affiliation{Theoretical Division, Los Alamos National Laboratory, Los Alamos, New Mexico 87545, USA}

\author{Avadh Saxena}
\affiliation{Theoretical Division, Los Alamos National Laboratory, Los Alamos, New Mexico 87545, USA}

\begin{abstract}
Recently spin textures called skyrmions have been discovered in certain chiral magnetic materials 
without spatial inversion symmetry, and have attracted enormous attention due to 
their promising application in spintronics since only a low applied current is necessary to drive their motion. When a conduction electron moves around the skyrmion, its spin is fully polarized by the spin texture  and acquires a quantized phase; thus, the skyrmion yields an emergent electrodynamics that in turn determines skyrmion motion and gives rise to a finite Hall angle. As topological excitations, skyrmions behave as particles. In this work we derive the equation of motion for skyrmions as rigid point particles from a microscopic continuum model and obtain the short-range interaction between skyrmions, and the interaction between skyrmions and defects. Skyrmions also experience a Magnus force perpendicular to their velocity due to the underlying emergent electromagnetic field. We validate the equation of motion by studying the depinning transition using both the particle and the continuum models. By using the particle description, we explain the recent experimental observations of the rotation of a skyrmion lattice in the presence of a temperature gradient. 
We also predict quantum and thermal creep motion of skyrmions in the pinning potential. 
 \end{abstract}
\date{\today}
\maketitle

\section{Introduction}
Spin texture called skyrmion was predicted to exist in certain magnetic materials. \cite{Bogdanov89,Bogdanov94,Rosler2006} Skyrmion  crystals have been recently observed in MnSi, $\rm{Fe_{0.5}Co_{0.5}Si}$, and other B20 transition metal compounds with small angle neutron scattering, 
Lorentz force microscopy and spin-polarized scanning tunneling microscopy. \cite{Muhlbauer2009,Munzer10,Pfleiderer10,Yu2010a,Yu2011,Heinze2011,Seki2012} These spin textures become more stable in thin films \cite{Yi09,Butenko2010,Yu2011,Heinze2011}, and they crystallize into a triangular lattice similar to that found for vortices in type II superconductors. 
The typical size of a skyrmion is about 10 nm and the corresponding lattice constant is about 100 nm. As more skyrmion crystals are discovered in new materials, it is expected that this state of matter will turn out to be a general form of magnetic ordering, existing ubiquitously in magnets without inversion symmetry.

A promising set of spintronics applications arises from the fact that skyrmions can be driven by a spin-polarized current as a result of the spin-transfer torque. The weak current required to move a skyrmion from the pinning center is 4 to 5 orders of magnitude smaller 
than the current required to move the well-studied magnetic domain  walls. \cite{Jonietz2010,Yu2012,Schulz2012} 
Therefore, skyrmions can be manipulated with much less energy dissipation.
A theoretical framework for understanding skyrmion dynamics is then crucial for applications.
Current descriptions are mostly based on continuum models that are difficult to solve analytically, and can be computationally intensive. Because skyrmions appear to have particle-like properties, the derivation  of a particle-based equation of motion provides a functional form for interactions between skyrmions, skyrmion-defect interactions, and the role of terms such
as the Magnus force.  Such a model would have tremendous impact on understanding skyrmion dynamics by theoretical analysis and computational modeling. The derivation of effective equations of motion for other systems, such as vortices in type-II superconductors~\cite{Blatter94}, has been crucial for understanding pinning and vortex dynamics  in the flux-flow regime. We note that the equation of motion for a single skyrmion has been reported recently. \cite{Zang11,Everschor11,Everschor12,Iwasaki2013,Liu2013,Liu2013b} However the full equation of motion including the interaction between skyrmions, and interaction between skyrmions and defects is not available. The purpose of the present work is to fill this gap and also to show several applications of the derived equation of motion.

In this article we derive a concise particle-like  equation of motion for skyrmions using the Thiele's approach \cite{Thiele72}. The emergent electromagnetism induced by the Berry phase leads to an additional Magnus force  that strongly suppresses the depinning current by deflecting skyrmions away from the pinning centers. By applying the derived equation of motion to the study of skyrmion lattice rotation in the presence of a temperature gradient, we reproduce recent experimental results.  We use the same equation of motion to investigate the quantum and thermal creep motion of a skyrmion in a pinning potential. Finally, we validate the particle model by computing depinning transitions and comparing against results obtained with the original continuum model.

\section{Equation of motion}
\begin{figure}[t]
\psfig{figure=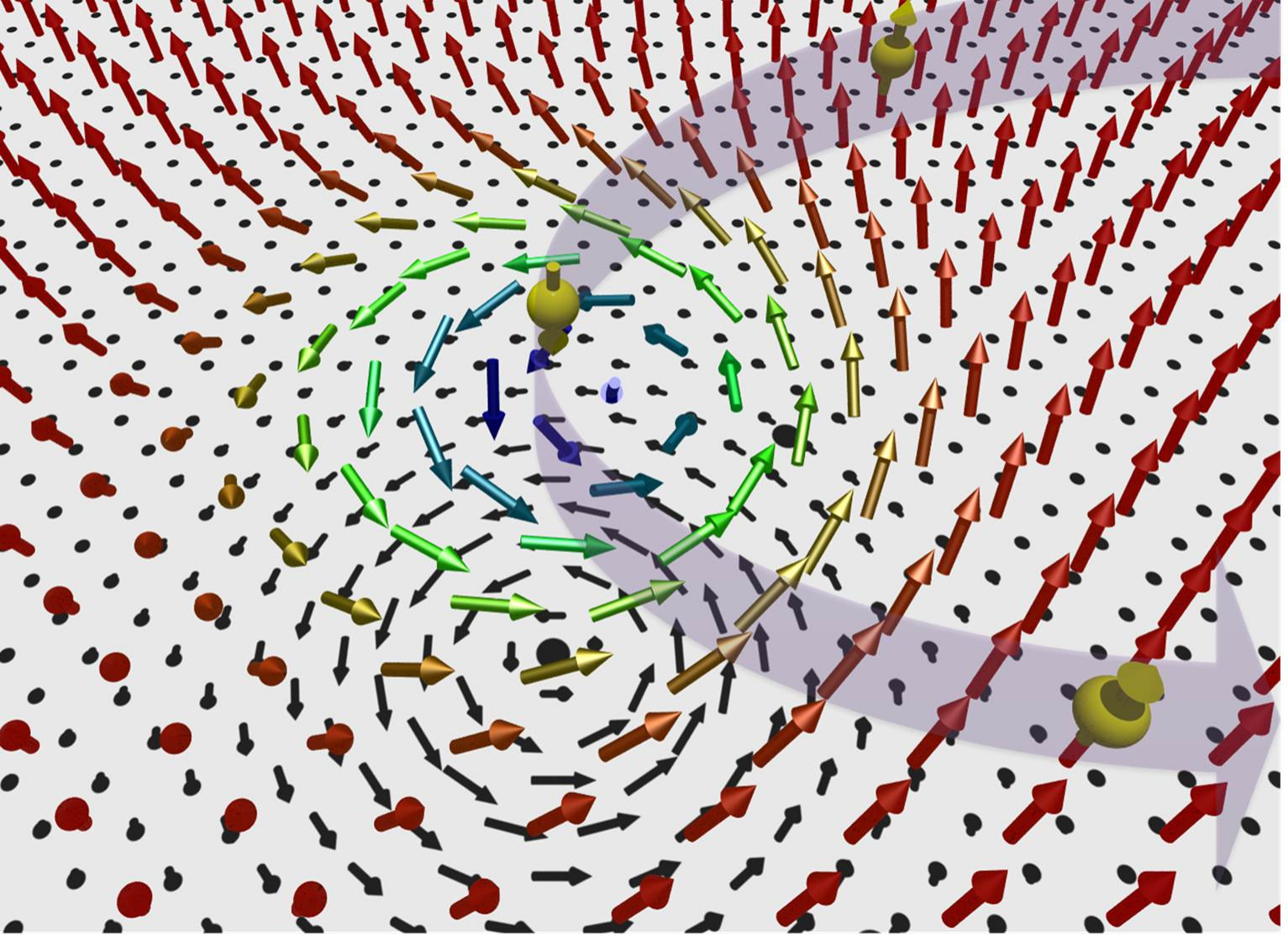,width=\columnwidth}
\caption{\label{f1}(color online) Schematic view of an electron (yellow sphere with arrow) passing through a skyrmion (colored arrows). The spin of the electron follows the spin texture of the skyrmion, giving rise to an emergent magnetic field that couples the electronic orbital motion. The black arrows (dots) are the spin projection in the $x-y$ plane and the spins in the ferromagnetic state are along the $z$ axis.}
\end{figure}

We consider a thin film of a chiral magnet with Dzyaloshinskii-Moriya (DM) interaction which supports skyrmions. \cite{Bogdanov89,Bogdanov94,Rosler2006,Han10,Rossler2011} The magnetic moments are described by a unit vector $\mathbf{n}(\mathbf{r})$. The corresponding action for the magnetic moments $\mathbf{n}$ can be written as
\begin{equation}\label{eq1}
S=S_B-\frac{d\alpha_g}{\gamma} \int dt dt' d^2r \left[\frac{\mathbf{n}(t)-\mathbf{n}(t')}{t-t'}\right]^2- \int dt \mathcal{H},
\end{equation}
\begin{equation}\label{eq2}
S_B=d\int d^2 r d t z^{\dagger} i\left( \frac{1}{\gamma}\partial_t-\frac{\hbar}{2 e} \mathbf{J}\cdot\nabla\right) z,
\end{equation}
where $z\equiv |z\rangle$ is the spin coherent state defined as $\mathbf{n}\cdot {\mathbf \sigma} |z\rangle= |z\rangle$. ${\mathbf \sigma}$ is the vector of Pauli matrices and $d$ is the film thickness. Here, $\gamma=a^3/(\hbar s)$ with $a$ the lattice constant and $s$ the total spin. The first term in $S_B$ describes the Berry phase for the precession of a spin at $\mathbf{r}=(x,\ y)$. In the presence of conducting electrons, the electrons become fully polarized by the local moments $\mathbf{n}$ in the large Hund's coupling limit, as depicted in Fig. \ref{f1}. The second term in $S_B$ is responsible for the Berry phase that the electron picks up when it moves around the skyrmion. The term proportional to $\alpha_g$ accounts for the Gilbert damping. The spin Hamiltonian is
\begin{equation}\label{eq3}
\mathcal{H}=d\int d\mathbf{r}^2 \left[\frac{J_{\rm{ex}}}{2}(\nabla \mathbf{n})^2+D\mathbf{n}\cdot\nabla\times \mathbf{n}-\mathbf{H}_a\cdot\mathbf{n} \right].
\end{equation}
The first term is the exchange interaction, the second term is the DM interaction, which breaks spatial inversion symmetry, and the last term is the Zeeman energy. The external magnetic field is perpendicular to the film. Systems governed by Eq. \eqref{eq3} support a skyrmion phase in an intermediate magnetic field $0.2D^2/J_{\rm{ex}}<H_a<0.8D^2/J_{\rm{ex}}$. \cite{Rossler2011} The skyrmion is characterized by the topological charge density $Q(\mathbf{r})=\int d r^2\mathbf{n}\cdot(\partial_x \mathbf{n}\times \partial_y \mathbf{n})/(4\pi)=\pm 1$. 

According to Eqs. \eqref{eq1} and \eqref{eq2}, the spin dynamics is governed by the Landau-Lifshitz-Gilbert equation \cite{Bazaliy98,Li04,Tatara2008}
\begin{equation}\label{eq4}
{\partial _t}{\bf{n}} = \frac{\hbar\gamma}{2e}({{\bf{J}} }\cdot\nabla) {\bf{n}} - \gamma {\bf{n}} \times {{\bf{H}}_{\rm{eff}}} + \alpha_g {\partial _t}{\bf{n}} \times {\bf{n}},
\end{equation}
with the effective magnetic field $\mathbf{H}_{\rm{eff}}\equiv\delta \mathcal{H}/\delta{\mathbf{n}}$. In metallic chiral magnets, the motion of skyrmions generates electric fields, hence induces a dissipative current $\mathbf{J}_{\mathrm{diss}}=\sigma \hbar [\mathbf{n}\cdot(\nabla\mathbf{n}\times\partial_t\mathbf{n})]/(2e)$, where $\sigma$ is the conductivity. \cite{Zang11} In insulating magnets, such a dissipative current is absent because $\sigma=0$. The current density in Eq. \eqref{eq4} thus is the sum of the external current $\mathbf{J}_{\mathrm{B}}$ and the dissipative current  $\mathbf{J}_{\mathrm{diss}}$, $\mathbf{J}=\mathbf{J}_{\mathrm{B}}+\mathbf{J}_{\mathrm{diss}}$.

Equations \eqref{eq1}-\eqref{eq4} describe the skyrmion dynamics as well as its deformations.  Skyrmions can be treated as particles as long as deformations of their internal structure remain small. In other words, a particle-like description assumes that skyrmions have a rigid internal structure. Such rigidity is determined by the frequency of the normal modes associated with small fluctuations of the spin texture around the stationary state. Here it is important to not that the Zeeman and the DM terms remove any continuous symmetry except for translations. Therefore, the only Goldstone mode arises from translations of the rigid skyrmion. Modes associated with the internal skyrmion structure always have a finite frequency. This finite frequency gap increases with magnetic field and provides a natural justification for treating skyrmions as particles.
To treat skyrmions as point particles, we make the following two approximations. {We assume a  skyrmion density such that the overlap between different skyrmions is small. We also} assume that the structure of moving skyrmions is the same as 
that in the static case. The internal structure of skyrmions becomes irrelevant under these conditions, which are satisfied in the low velocity region for certain magnetic fields.~\cite{szl13a}  We first derive the equation of motion for a skyrmion in the particle-level description based on the Thiele's approach \cite{Thiele72}:
\begin{equation}\label{eq5}
\frac{4\pi\alpha  }{\gamma}\mathbf{v}_i=\mathbf{F}_M+\mathbf{F}_L+\sum_j\mathbf{F}_d(\mathbf{r}_j-\mathbf{r}_i)+\sum_j \mathbf{F}_{ss}(\mathbf{r}_j-\mathbf{r}_i),
\end{equation}
where $\mathbf{v}_i$ is the skyrmion velocity.  Equation ~\eqref{eq5} is the main result of the present work. The term on the left-hand side accounts for the damping of skyrmion motion, which is produced by the underlying damping of the spin precession and damping due to the conduction electrons localized in the skyrmions. Thus $\alpha=\alpha_g\eta+\alpha_{\sigma}\eta'$ with
\begin{equation}\label{eq5aa}
\alpha_{\sigma}=4\pi\left(\frac{\hbar}{2e\xi_s}\right)^2\gamma\sigma,
\end{equation}
\begin{equation}\label{eq5ab}
\eta=\eta_{\mu}=\frac{1}{4\pi}{\int_{\mathrm{skyrmion}} {{d}}{\mathbf{r}^2}{({\partial _{\mu}}{\bf{n}})^2}},
\end{equation}
\begin{equation}\label{eq5ac}
\eta'=\frac{\xi_s^2}{16\pi^2}\int_{\mathrm{skyrmion}} d r^2[\mathbf{n}\cdot(\partial_x\mathbf{n}\times\partial_y\mathbf{n})]^2.
\end{equation}
where the integration in Eqs. \eqref{eq5ab} and Eqs. \eqref{eq5ac} is performed around the skyrmions and $\xi_s\sim J_{\mathrm{ex}}/D$ is the size of skyrmions. Here $\mu=x,\ y$ in Eq. \eqref{eq5ab}. $\eta\approx 1$ and $\eta'\approx 1$ depends weakly on $J_{\rm{ex}}/D$ for $J_{\rm{ex}}/D\gg a$. For typical parameters, $\alpha_\sigma\gg \alpha_g$. \cite{Zang11} In Eq. \eqref{eq5}, $\mathbf{F}_M=4\pi\gamma^{-1}\hat{z}\times \mathbf{v}_i$ is the Magnus force per unit length, which is perpendicular to the velocity. $\mathbf{F}_L=2\pi\hbar e^{-1}\hat{z}\times \mathbf{J}_{\mathrm{B}}$ is the Lorentz force due to the external current, that arises from the emergent quantized magnetic flux $\Phi_0=hc/e$ carried by the skyrmion in the presence of a finite current. $\mathbf{F}_{ss}$ is the pairwise interaction between two skyrmions and $\mathbf{F}_d$ is the interaction between skyrmions and quenched disorder. It is clear from Eq. \eqref{eq5}, that the rigid skyrmion does not have an intrinsic mass. For thin films, the skyrmions are straight in the direction perpendicular to the film, and the forces in Eq. \eqref{eq5} are defined per unit length. A similar equation of motion was considered before in the context of vortices of type II superconductors.~\cite{Blatter94} However, the Magnus force is negligibly small for superconducting vortices in most cases.~\cite{Blatter94}

The equation motion for a single skyrmion [without the terms $\mathbf{F}_d$ and $\mathbf{F}_{ss}$ in Eq. \eqref{eq5}] can be derived heuristically. In stationary state, the skyrmion and conduction electrons form a composite object and move together. Let us focus on the conduction electrons inside the skyrmion. The electric current density inside the skyrmion is
\begin{equation}\label{eq5ad}
\mathbf{J}_{{e}}=\sigma_{\parallel} \mathbf{E}+\sigma_{\perp}\hat{z}\times\mathbf{E},
\end{equation}
where $\sigma_{\parallel}$ and $\sigma_{\perp}$ is the longitudinal and Hall conductivity. Using the Drude model, we have
\begin{equation}\label{eq5ae}
\sigma_{\parallel}=\frac{e^2\rho_e\tau_e}{m}\frac{1}{1+(\omega_c\tau_e)^2}, \ \ \sigma_{\perp}=\frac{e^2\rho_e\tau_e}{m}\frac{\omega_c\tau_e}{1+(\omega_c\tau_e)^2},
\end{equation}
where $\tau_e$ is the electron relaxation time, $\rho_e\sim 1/a^3$ is the electron density and $\omega_c=e B_{e}/(m_ec)$ is the cyclotron frequency with the electron mass $m_e$, because the electrons experience the emergent magnetic field $B_e\approx \Phi_0/(\xi_s^2)$. The electric field is $\mathbf{E}=\mathbf{B}_e\times \mathbf{v}/c$. Substituting $\mathbf{E}$ into Eq. \eqref{eq5ad} and taking the cross product $\times \Phi_0\hat{z}/c$ at both sides of Eq. \eqref{eq5ad}, we obtain the equation motion for conduction electrons
\begin{equation}\label{eq5af}
\pi\rho_e\frac{\omega_c\tau_e}{1+(\omega_c\tau_e)^2}\mathbf{v}=\pi\rho_e\frac{(\omega_c\tau_e)^2}{1+(\omega_c\tau_e)^2}\hat{z}\times\mathbf{v}+\hat{z}\times\mathbf{J}_e\Phi_0/c,
\end{equation} 
which is also the equation of motion for the skyrmion. For a strong internal field, $\omega_c\tau\gg 1$ and the Magnus force dominates. The derivation based on the Landau-Lifshitz-Gilbert equation is present in Sec. II A.

The action for the particle model of Eq. \eqref{eq5} can be written as
\begin{equation}\label{eq6}
\frac{S_p}{d}=S_{B,p}-U(\mathbf{r})-\frac{4\pi\alpha }{\gamma}\int dt dt' \left[\frac{\mathbf{r} (t)-\mathbf{r} (t')}{t-t'}\right]^2,
\end{equation}
\begin{equation}\label{eq7}
S_{B,p}=\frac{4\pi}{\gamma} \left[x \left(\frac{1}{2}\partial_t y- \frac{\hbar \gamma}{2e}J_y\right)-y \left(\frac{1}{2}\partial_t x- \frac{\hbar \gamma}{2e}J_x\right)\right],
\end{equation}
where $U(\mathbf{r})$ is the potential per unit length produced by other skyrmions and pinning sites. A unique feature is that $x$ and $y$ are conjugate variables, which is a hallmark of the Berry phase, as given by $S_B$ in Eq. \eqref{eq2}.

One can treat the Magnus force $\mathbf{F}_M$ as originating from an effective transverse magnetic field, $B_z=4\pi c d/(\gamma q)$, that couples to a charged moving particle with charge $q$. This emergent magnetic field originates from the Berry phase. The Magnus force does not produce work, but it affects the skyrmion trajectory. As we will see later,  skyrmions are easily deflected by  pinning centers  because of the Magnus force. This effect explains the very weak pinning that has been observed in different experiments.

In the present work, we focus on the adiabatic spin transfer torque described by the term, $\hbar\gamma(\mathbf{J}\cdot\nabla)\mathbf{n}/(2e)$, in Eq. \eqref{eq4}. Generally, there will be also non-adiabatic spin transfer torque given by the expression,  $-\zeta\hbar\gamma\mathbf{n}\times(\mathbf{J}\cdot\nabla)\mathbf{n}/(2e)$. \cite{Tatara2008} Our derivation is readily generalized to the non-adiabatic spin transfer torque, which yields additional force at the right-hand side of Eq. \eqref{eq5}, $\mathbf{F}_{\mathrm{non}}=2\pi\hbar\zeta\eta e^{-1}\mathbf{J}_{\mathrm{B}}$
\begin{equation}
\frac{4\pi\alpha  }{\gamma}\mathbf{v}_i=\mathbf{F}_M+\mathbf{F}_L+\mathbf{F}_{\mathrm{non}}+\sum_j\mathbf{F}_d(\mathbf{r}_j-\mathbf{r}_i)+\sum_j \mathbf{F}_{ss}(\mathbf{r}_j-\mathbf{r}_i).
\end{equation}
The effects of the non-adiabatic spin transfer torque on skyrmion dynamics was studied recently in Ref. \onlinecite{Iwasaki2013}.

We remark that Eq. \eqref{eq5} can be readily generalized to the case with skyrmions in insulating magnets. In this case, the damping due to the conduction electrons is absent and $\alpha=\alpha_g\eta$. Both the Lorentz force $\mathbf{F}_L$ and the force due to the non-adiabatic spin transfer torque $\mathbf{F}_{\mathrm{non}}$ are absent.

We estimate the force using the typical parameters for MnSi, \cite{Zang11} $a\approx 2.9\ \AA$, $J_{\rm{ex}}\approx 3\ {\rm{meV}}/a$, $D\approx 0.3\ {\rm{meV}}/a^2$, $\alpha \approx 0.1$ and $s\approx 1$. At a velocity $v=1\ \rm{m/s}$, we estimate the dissipative force per unit length to be $F_{\rm{diss}}\equiv 4\pi\alpha v/(\gamma)\approx 5\times 10^{-6}\ \rm{N/m}$; the Magnus force per unit length is $F_M\approx 5\times 10^{-5}\ \rm{N/m}$. Thus $F_M \gg F_{\rm{diss}}$. The repulsive force per unit length between skyrmions for $d\approx 20\ \rm{nm}$ is $F_{ss}\approx 10^{-5}\ \rm{N/m}$ at a separation $r_d=10\ \rm{nm}$ (see Fig. \ref{f3} below). The Lorentz force per unit length at a current density $J_{\mathrm{B}}=10^6\ \rm{A/m^2}$ is $F_L\approx 4\times 10^{-9}\ \rm{N/m}$. Since the depinning current for skyrmion lattice is of the order of $10^6\ \rm{A/m^2}$, \cite{Jonietz2010,Yu2012,Schulz2012} we thus estimate the pinning fore per unit length as $F_d\approx 4\times 10^{-9}\ \rm{N/m}$. 

\begin{figure}[t]
\psfig{figure=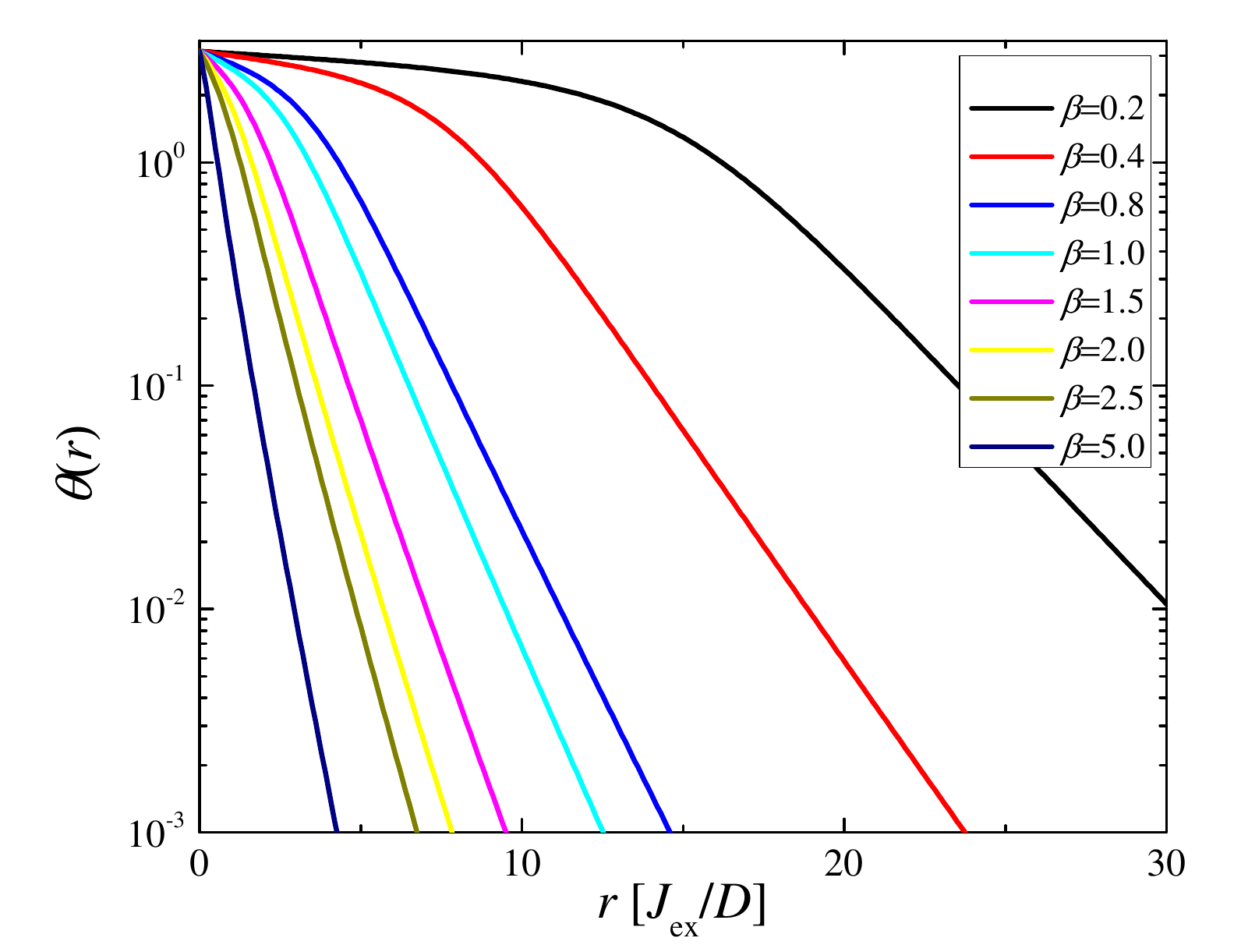,width=\columnwidth}
\caption{\label{f2}(color online) Profile of $\theta(r)$ obtained from a numerical solution of Eq. \eqref{eq10} for different values of the magnetic field.}
\end{figure}
\begin{figure}[b]
\psfig{figure=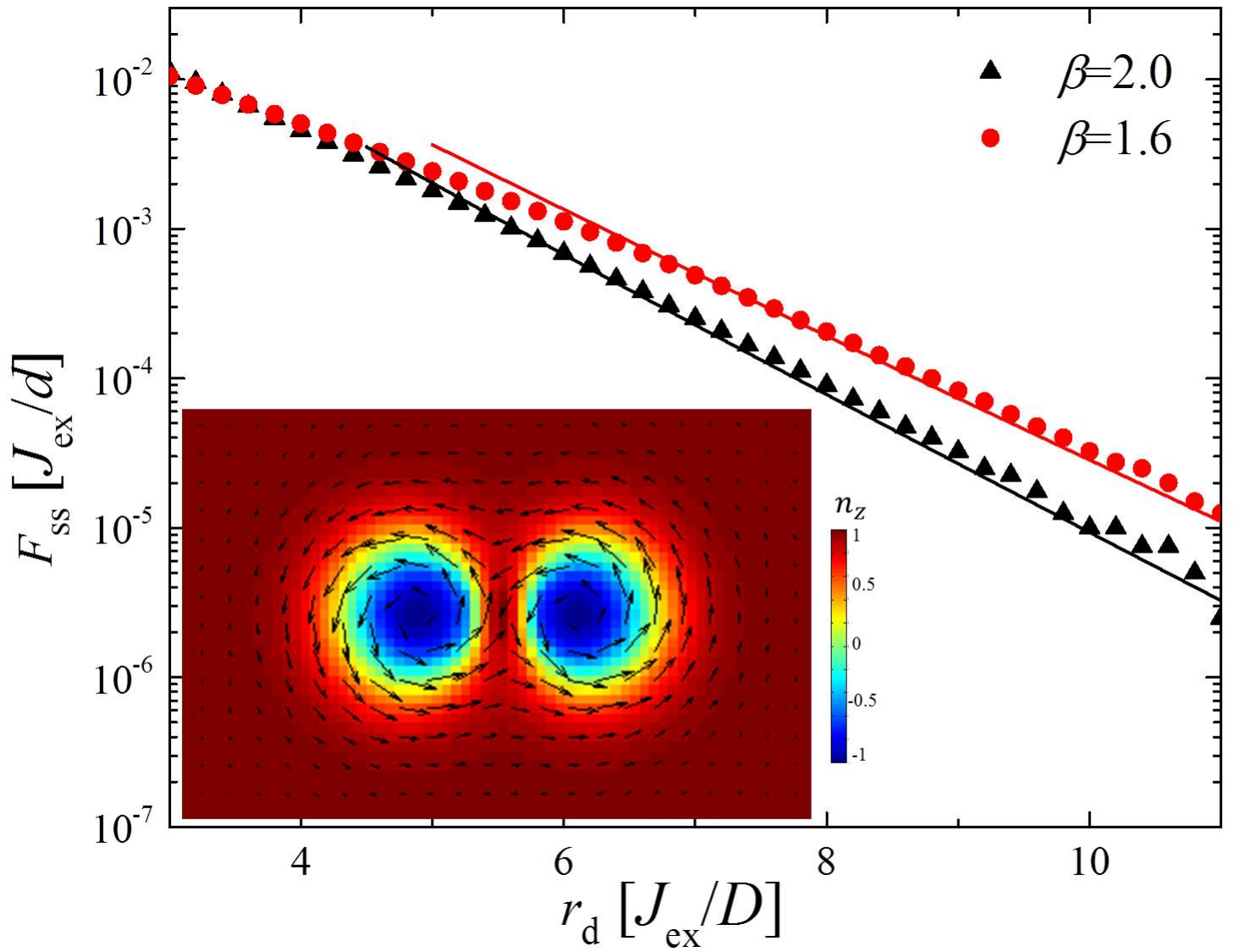,width=\columnwidth}
\caption{\label{f3}(color online) Force between two skyrmions as a function of the separation $r_d$ in two different magnetic fields. Symbols are obtained from a numerical solution of Eq. \eqref{eq4} and lines are fits to $K_1(r_d/\xi)$, with $\xi=D/\sqrt{H_a J_{\rm{ex}}}$. Inset:  stationary configuration of two skyrmions at $r_d=3.6J_{\rm{ex}}/D$. The vectors denote the $n_x$ and $n_y$ components and the $n_z$ component is represented by the color scale.}
\end{figure}

\subsection{Derivations}
Spins precess collectively when a rigid skyrmion moves with velocity $\mathbf{v}$, $\mathbf{n}_s(\mathbf{r}-\mathbf{v} t)$, and their evolution is governed by the equation of motion 
\begin{equation}\label{eq8}
\partial_t \mathbf{n}_s=\frac{\hbar\gamma}{2e}(\mathbf{J}\cdot\nabla)\mathbf{n}_s-\gamma\mathbf{n}_s\times \mathbf{H}_i+\alpha_g \mathbf{n}_s\times\partial_t \mathbf{n}_s,
\end{equation}
where $\mathbf{H}_i=\mathbf{H}_s+\mathbf{H}_d$. $\mathbf{H}_s$ is the magnetic field produced by other skyrmions and $\mathbf{H}_d$ is the field produced by defects. The effective field $\mathbf{H}_0\equiv \delta\mathcal{H}/\delta\mathbf{n}_s$ due to the skyrmion $\mathbf{n}_s$ does not contribute to $\mathbf{H}_i$ because $\mathbf{n}_s\times\mathbf{H}_0=0$ for a rigid skyrmion.  If we first multiply both sides of Eq. (\ref{eq8}) by $\times \mathbf{n_s}$ (cross product) and then by $\cdot\partial_\mu \mathbf{n}_s$ (dot product), we obtain 
\begin{equation}\label{eq9}
{\alpha} \mathbf{v}=\frac{\gamma}{4\pi}\left[\mathbf{F}_M+\mathbf{F}_L+\int dr^2 \mathbf{H}_{\perp}(\mathbf{r}'-\mathbf{r})\cdot\nabla_{r} \mathbf{n}_s(\mathbf{r}) \right],
\end{equation}
after integrating over the area around the skyrmion. Here $\mathbf{H}_{\perp}$ is the field component perpendicular to $\mathbf{n}_s$ and we have used that $\int dr^2 \partial_x \mathbf{n}_s\cdot\partial_y \mathbf{n}_s=0$ for a rigid skyrmion. The interaction potential between a skyrmion at $\mathbf{r}$ and an another skyrmion at $\mathbf{r}'$ is $U_{ss}(\mathbf{r}'-\mathbf{r})=-\int d r''^2\mathbf{n}_s(\mathbf{r}-\mathbf{r}'') \cdot \mathbf{H}_{s}(\mathbf{r}'-\mathbf{r}'')$ and the corresponding force is  $\mathbf{F}_{ss}=\int d r''^2\nabla_r \mathbf{n}_s(\mathbf{r}-\mathbf{r}'')\cdot\mathbf{H}_{s, \perp}(\mathbf{r'}-\mathbf{r}'')$. The self-energy of the skyrmion in the presence of defects is $E_s(\mathbf{r}-\mathbf{r}')=-\int dr''^2\mathbf{n}_s(\mathbf{r}-\mathbf{r}'')\cdot \mathbf{H}_d(\mathbf{r}'-\mathbf{r}'')$, where $\mathbf{H}_d(\mathbf{r})=J_{\rm{ex}}(\mathbf{r})\nabla^2 \mathbf{n}_s/2-D(\mathbf{r})\nabla\times \mathbf{n}_s+\mathbf{B}$. The pinning force is then given by $\mathbf{F}_d=\int dr''^2\nabla \mathbf{n}_s(\mathbf{r}-\mathbf{r}'')\cdot \mathbf{H}_{d, \perp}(\mathbf{r'}-\mathbf{r}'')$. Thus, Eq.~\eqref{eq9} reduces to Eq. \eqref{eq4} if we replace the integral by the interaction force.

To calculate the interaction between skyrmions and the interaction between skyrmions and defects, we need to know the structure of a single skyrmion. An isolated skyrmion is described by $\mathbf{n}_s(r, \phi)= \sin\theta\hat{\phi}+\cos\theta\hat{z}$ in the polar coordinates $(r, \phi)$ with $\hat{\phi}$ and $\hat{z}$ being the unit vectors along the corresponding axises.  $\theta(r)$ is determined by minimizing $\mathcal{H}$ in Eq.~\eqref{eq3},
\begin{equation}\label{eq10}
-r\partial _r^2\theta-\partial_r\theta+\cos(2\theta) +\frac{\sin (2\theta)}{2 r}+\frac{\beta}{2} r \sin(\theta) -1=0,
\end{equation}
with the boundary condition $\theta(r=0)=\pi$ and $\theta(r\rightarrow +\infty)=0$, where we have renormalized the distance $r$ as $r \to r/(J_{\rm{ex}}/D)$, and $\beta={2 H_a J_{\rm{ex}}}/{D^2}$. The profile of $\theta(r)$ for different $\beta$ is shown in Fig. \ref{eq2}. There are two length scales associated with a skyrmion.  $\theta$ decreases linearly in $r$ for $r\ll 1$, while the asymptotic solution far away from the center of the skyrmion, $r\rightarrow \infty$, is $\theta\sim K_0(r/\xi)$ with a healing length $\xi=\sqrt{{2}/{\beta}}$. Here $K_0$ and $K_1$ below are the modified Bessel functions. The spin recovers exponentially to the fully polarized state due to the finite energy gap in the spectrum of the spin wave excitations  that is induced by the external field. One may define the core region of the skyrmion as $\theta(r<R_c)<\pi/2$.

\begin{figure}[t]
\psfig{figure=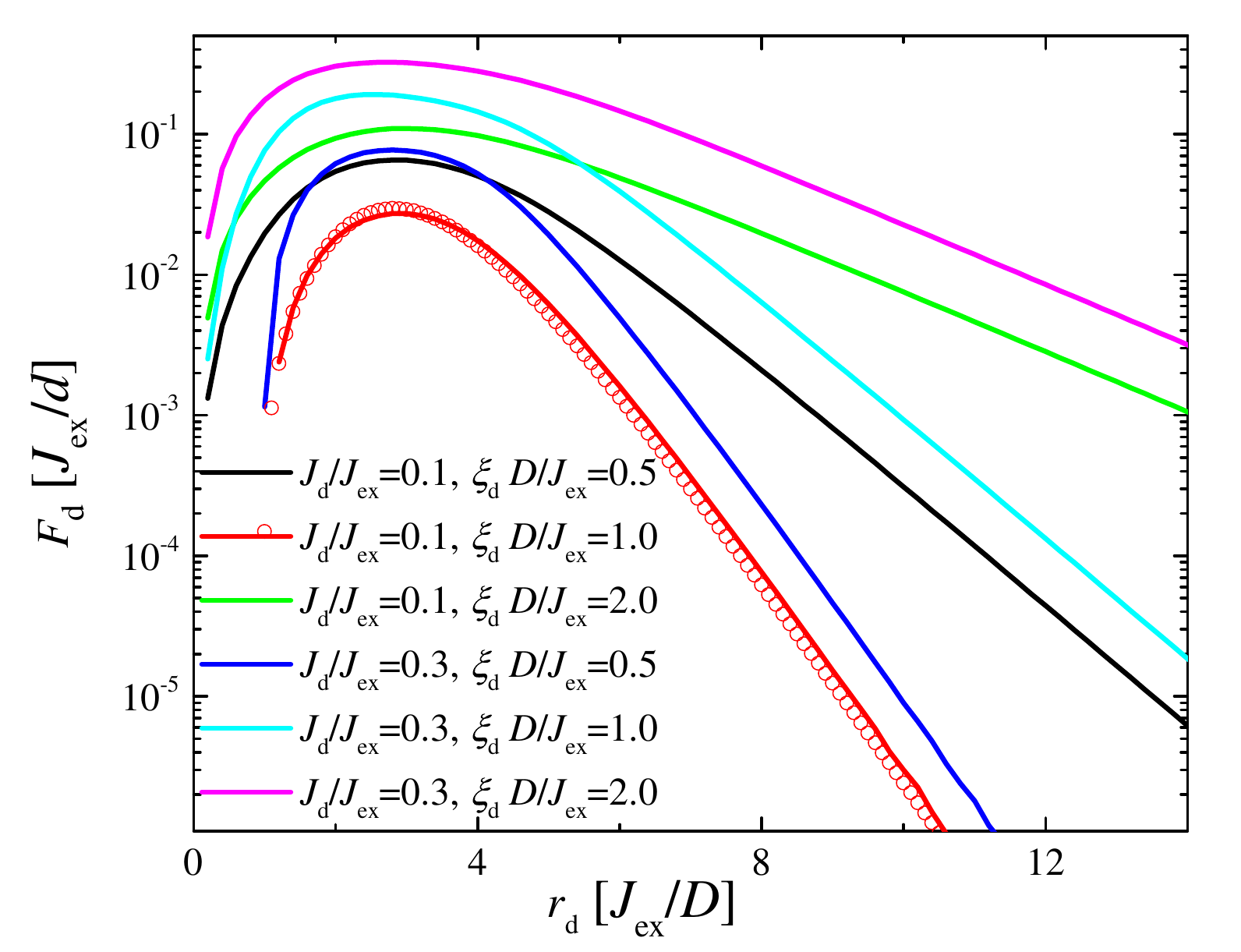,width=\columnwidth}
\caption{\label{f4}(color online) (a) Interaction force between a skyrmion and a defect for different strength and size of defect. Lines are results obtained from full numerical calculations and symbols are results obtained by assuming a rigid skyrmion structure.}
\end{figure}

Since the interaction between two skyrmions is induced by the overlap between both spin textures, it must depend on the length $\xi$. To calculate the interaction between two skyrmions numerically (see Appendix A), we initially pinned the two skyrmions at a fixed separation, $r_d$, by freezing the spins within a radius $r\le J_{\rm{ex}}/D$ during the time evolution dictated by Eq. \eqref{eq4}, and calculated the energy as a function $r_d$. The results shown in Fig.~\ref{f3} indicate that the interaction decays exponentially and it is well described by $F_{ss}\sim K_1(r_d/\xi)$.

We next proceed to study the interaction between skyrmions and defects. The electronic density is not homogeneous in real systems leading to an inhomogeneous exchange interaction $J_{\rm{ex}}$ produced by the double-exchange mechanism. We model the defects by the following profile of $J_{\rm{ex}}$:
\begin{equation}\label{eq11}
J_{\rm{ex}}(r)=J_0\left(1+\sum_iJ_d\exp[-|\mathbf{r}-\mathbf{r}_{d,i}|/\xi_d]\right),
\end{equation}
where $J_d$ characterizes the strength of the defects and $\mathbf{r}_{d,i}$ is the pinning center. The characteristic size of the defects, $\xi_d$, is comparable to the inter-atomic separation. For weak pinning, we can still use the rigid approximation for skyrmions. In this case the interaction energy is just the self-energy of the skyrmion given by using $J_{\rm{ex}}(r)$ in Eq. \eqref{eq11}. We first obtain the structure of the skyrmion from Eq. \eqref{eq10} and then calculate its self-energy with Eq. \eqref{eq3}. We also do a full numerical relaxation by holding the spin at the center of the skyrmion unchanged in order to pin the skyrmion at a desired position. Both methods yield consistent results, as shown in Fig. \ref{f4}. Several observations are as follows: 1) The length scale of the exponentially decaying force at a large distance is given by the size of defects $\xi_d$. Because the main contribution to the self-energy comes from the core of the skyrmion, the exponential tail does not contribute significantly and the interaction range is determined by $\xi_d$. 2) The force is maximized when the separation becomes close to the skyrmion radius, $r_d\approx R_c$, and it drops when the skyrmion gets even closer to the defect and finally vanishes when the core coincides with the center of the defect. 3)  Because the amplitude of the force is proportional to the strength of the defects, the force can be expressed as $F_d\sim J_d {\exp  \left(-r_d/\xi _d\right)}$ for large separations. 4) The force is attractive for $J_d<0$ and  it is repulsive for $J_d>0$, so skyrmions prefer to stay in the $J_d<0$ region to minimize their self-energy. 5) The non-uniformity of electron density  in real solids  (size of defects) is of the order of the inter-atomic length, $\xi_d\sim 0.1$ nm, that is much smaller than the typical skyrmion size. Thus, the interaction between defects and skyrmions is extremely weak. This is one of the reasons why the pinning of skyrmions is very weak.

\begin{figure}[b]
\psfig{figure=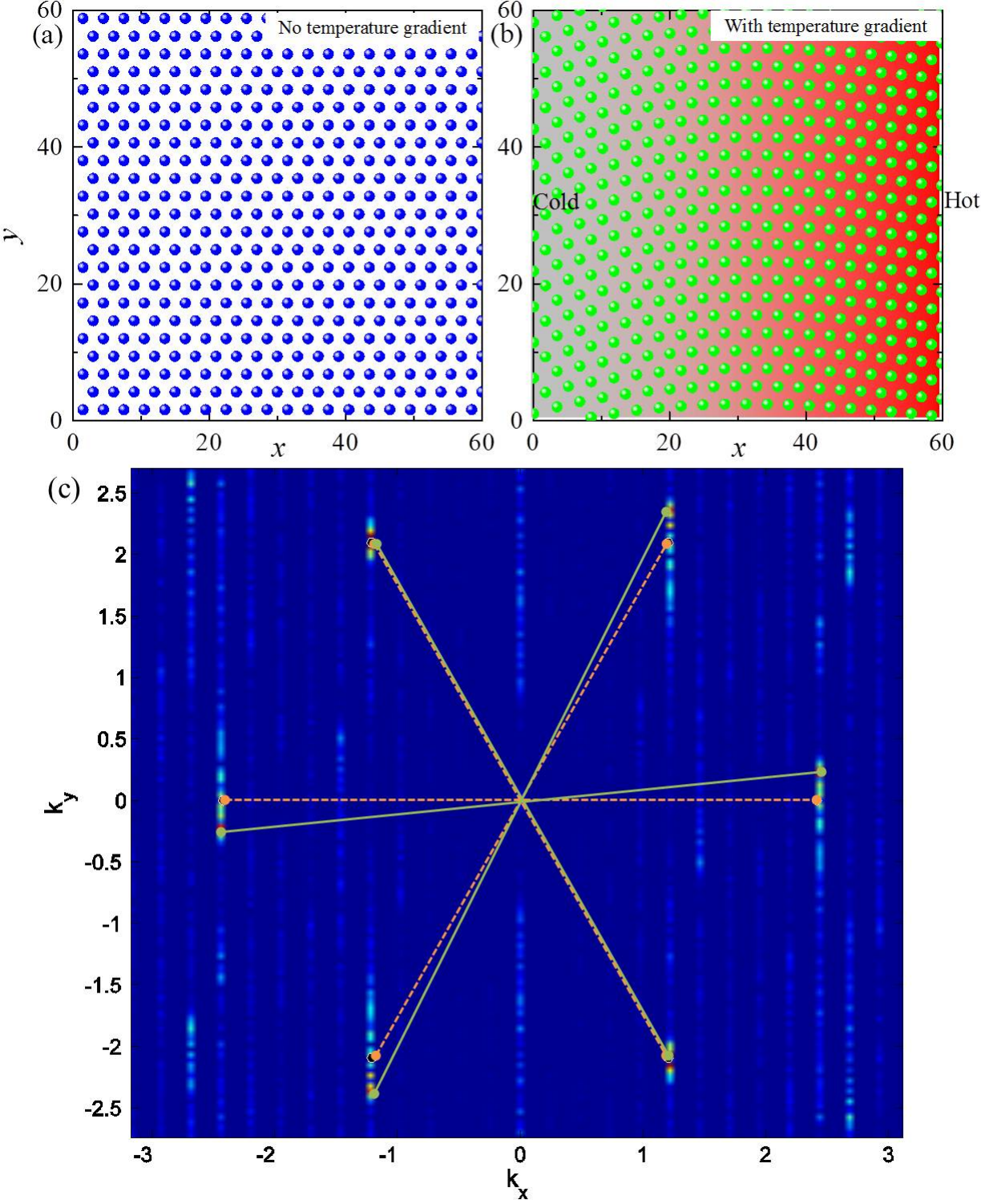,width=\columnwidth}
\caption{\label{f5}(color online) (a) and (b) Real-space configuration of the skyrmion lattice in the absence of temperature gradient (a) and in the presence of temperature gradient (b). For clarity, only part of the configuration is shown. (c) Bragg peaks (green dots and lines) of the skyrmion configurations with a temperature gradient. The triangular skyrmion lattice is rotated and distorted. The Bragg peaks for a perfect triangular lattice without a temperature gradient (orange lines and dots) are shown for comparison.}
\end{figure}
\section{Applications}

We apply the particle model of skyrmions to study the rotation of the skyrmion lattice in the presence of a temperature gradient, as well as the creep motion. We also compare the particle and continuum models by considering the depinning transition.

\subsection{Rotation of skyrmion lattice}
Recent neutron scattering experiments have shown that the skyrmion lattice rotates in the presence of a temperature gradient. \cite{Jonietz2010} The rotation was explained  in Refs.~\onlinecite{Everschor11,Everschor12} by using a continuum model. Here we show that the rotation can also be explained by using our particle model in Eq. \eqref{eq5}.

$\mathbf{F}_{ss}=\mathbf{F}_d=0$ in the crystal phase without defects. The Hall angle of the skyrmion trajectory is $\tan \theta_{H}=v_y/v_x=-\alpha$ when the current is along the $x$ direction. The damping coefficient, $\alpha$, has two contributions: the Gilbert damping and the dissipation due to the electric field induced by the skyrmion motion \cite{Zang11}. The latter contribution is dominant. Thus, $\alpha\sim \sigma$, where $\sigma(T)$ is the temperature-dependent conductivity. The temperature gradient leads to a spatial variation of the Hall angle that exerts a finite torque on the skyrmion lattice. In the absence of pinning, the lattice keeps rotating with a finite angular velocity. However, pinning is always  present real systems due to the underlying atomic crystal structure that favors one particular orientation of the skyrmion crystal. An additional pinning arises from the geometric confinement of finite samples.  The competition between  torque and crystal pinning thus yields to a stationary state in which  the skyrmion lattice is rotated by a finite angle. The lattice keeps rotating with finite angular velocity for a sufficiently large torque induced by large enough currents and/or temperature gradients.

We perform numerical simulations of Eq. \eqref{eq5} by modeling the temperature gradient with $\alpha(x)=0.6-0.5(2x-L_x)^2/L_x^2$, where $L_x$ is the length along the $x$ direction. The current is also parallel to the $x$ direction and we use a simulation box with aspect ratio $L_x:L_y=2:\sqrt{3}$ and periodic boundary conditions. These boundary conditions favor a particular orientation of the principal axis of the skyrmion lattice, which is parallel to the $x$ direction. In the stationary state, we find that the skyrmion lattice is rotated by a finite angle relative to the case of zero temperature gradient $\alpha_0=0.1$ [see Fig. \ref{f5}]. In addition to the rotation, there is a small distortion of the skyrmion lattice. Thus the experimental observation can be explained with the particle model in Eq.~\eqref{eq5}.

\subsection{Creep motion of skyrmions}
The skyrmions can easily leave the pinning potential either by quantum or/and thermal fluctuations because pinning is weak. This phenomenon leads to creep motion. We consider the dynamics of a single skyrmion in a pinning potential $U(r)$. Because $\alpha\ll 1$ for real materials, such as MnSi, we will neglect the damping for simplicity. The quantum creep rate for Eq. \eqref{eq5} was calculated in Refs. \onlinecite{Feigelman1993,Blatter94} for superconducting  vortices. To be specific, we will consider a pinning potential per unit length
\begin{equation}\label{eq12}
U(x, y)=U_d\left(\frac{y^2}{\lambda^2}+\frac{x^2}{\lambda^2}-\frac{x^3}{\lambda^3}\right).
\end{equation}
The action of Eq.~\eqref{eq6} becomes
\begin{equation}\label{eq13}
\frac{\mathcal{S}_p}{d}=\int d\tau\left[\frac{4\pi}{\gamma}y {\partial_\tau x}-\frac{y^2 U_d}{\lambda^2}+U_d \left(\frac{x^2}{\lambda^2}-\frac{x^3}{\lambda^3}\right)\right],
\end{equation}
in the imaginary time representation $t\rightarrow -i\tau$. Here $y$ plays the role of momentum; thus, if the potential is separable, i.e. $U(x, y)=U_1(x)+U_2(y)$, the $y$ dependent potential is not inverted in the imaginary time representation. Equation ~\eqref{eq13} is the same as the one for a particle with mass $\tilde{m}=\lambda^2/(2U_d)$ moving in a one dimensional potential $U(x)/U_d=x^2/\lambda^2-x^3/\lambda^3$. The skyrmion does not have an intrinsic mass according to Eq.~\eqref{eq5}. However, it gains an extrinsic mass in the presence of a  pinning potential. The quantum rate, $\Gamma_q\sim \exp(-\mathcal{S}_q/\hbar)$, with
\begin{equation}\label{eq14}
\frac{\mathcal{S}_q}{\hbar }=\frac{32\pi \lambda ^2 d}{15 \gamma \hbar },
\end{equation}
is independent of the height of the pinning potential, but it depends on the width. The quantum tunneling of skyrmions between pins is weak because $\mathcal{S}_q/\hbar\sim \lambda^2 d/a^3\gg 1$.

We now consider the escape rate due to thermal fluctuations. For this purpose we add a noise force $F_n$ in Eq. \eqref{eq5}
which satisfies 
\begin{equation}
\left\langle F_n\right\rangle=0, \ \ \left\langle F_n(t)F_n(t')\right\rangle=2k_B T \frac{4\pi\alpha}{\gamma} \delta(t-t'),
\end{equation}
according to the fluctuation-dissipation theorem.
The thermal rate, $\Gamma_T= \Omega \exp(-\Delta U/k_B T) $, is dominated by the exponential factor, $\exp(-\Delta U/k_B T)$, where $\Delta U=4 U_d d/27$ is the height of the pinning potential. This factor reflects the Boltzmann distribution of the skyrmion in the potential $U$, and is thus independent of the dynamics (see Appendix B). In contrast, the attempt frequency, $\Omega$, does depend on the dynamics and the Magnus force.

Thermal escape becomes dominant at high temperatures, while quantum creep is dominant in the low temperature region. The crossover temperature between quantum and thermal tunneling  is $k_B T^*=\Delta U\hbar /\mathcal{S}_q=5\gamma\hbar U_d/(72\pi\lambda^2)$, which depends on the ratio of the width of the pinning potential and its height. $T^*$ can be increased for  properly engineered pinning potentials.  Since the skyrmion carries a magnetization that is opposite to the ferromagnetic background, the creep motion  manifests itself in experiments as a decay of the opposite magnetization. Thus, like in the case of superconducting vortices, the rate can be obtained by measuring  the time dependence of the magnetization.

\begin{figure}[t]
\psfig{figure=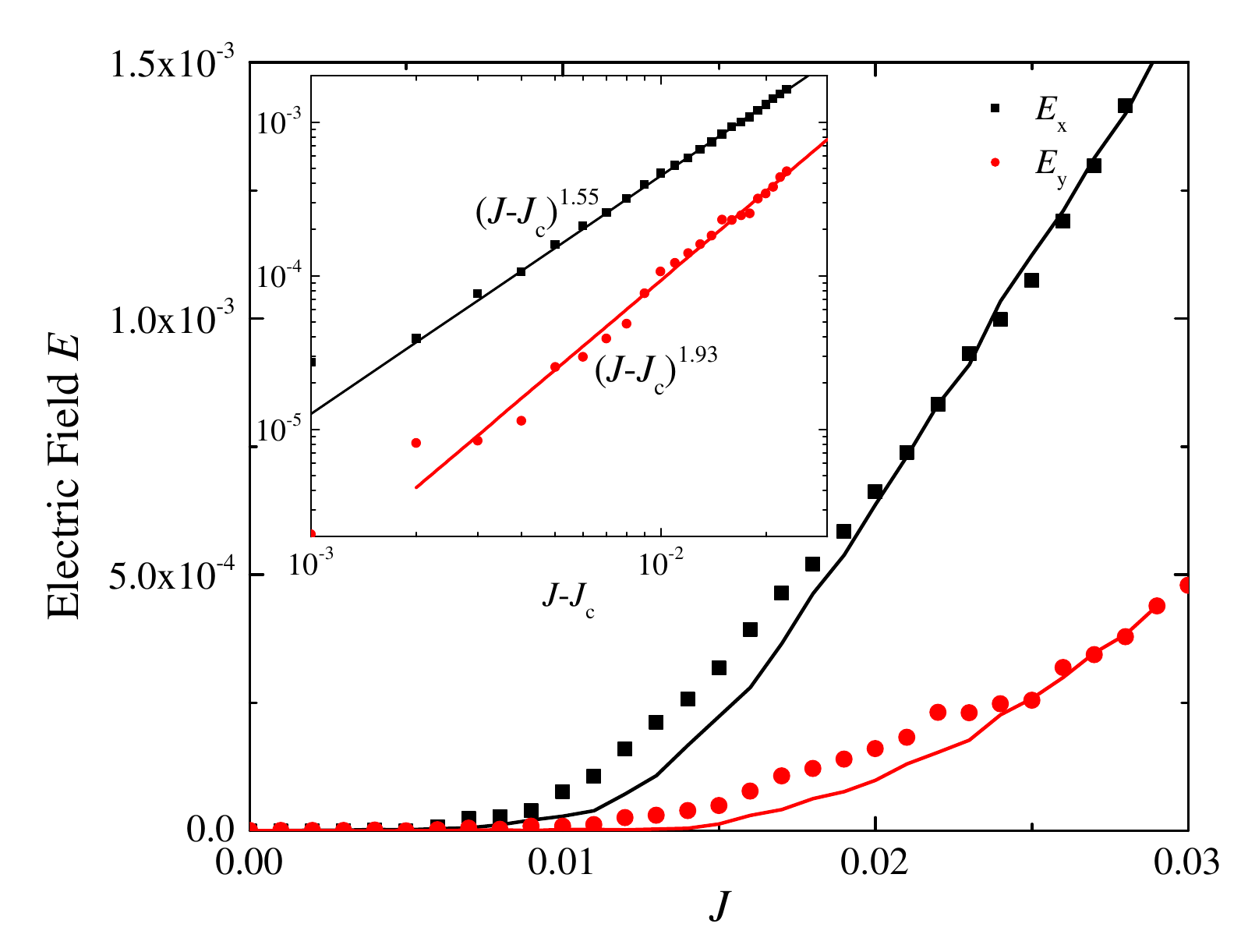,width=\columnwidth}
\caption{\label{f6}(color online) Comparison between the continuum and particle models. The lines are obtained with the continuum model while the symbols are obtained with the particle model. We convert the velocity of skyrmions in the particle model to an electric field by multiplying by an appropriate scaling factor. The inset is the scaling of the electric field near the depinning transition with $J_c=0.007$. The lines in the inset are power-law fits.}
\end{figure}

\subsection{Comparison between continuum and particle models}
To validate the particle model, we perform numerical simulations with both the particle and continuum models. We calculate the velocity of skyrmions as a function of the driving force when defects are present. The defects are modeled as in Eq. \eqref{eq11}. Skyrmions are pinned in a low driving current, and they depin from the defects when the Lorentz force is high enough. The particle model yields results that are in reasonable agreement with the continuum model as shown in Fig. \ref{f6}. Near depinning, the electric field behaves as $E_\mu\sim (J-J_c)^{\beta_\mu}$, where $J_c$ is the depinning current. From the numerical data we obtain $\beta_x\approx 1.55$ and $\beta_y\approx 1.93$. The exponent $\beta_\mu>1$ indicates that the depinning is plastic, i.e., some skyrmions escape from the pinning centers, while the others remain pinned. Eventually, all skyrmions become depinned when the current is further increased. 

\begin{figure}[t]
\psfig{figure=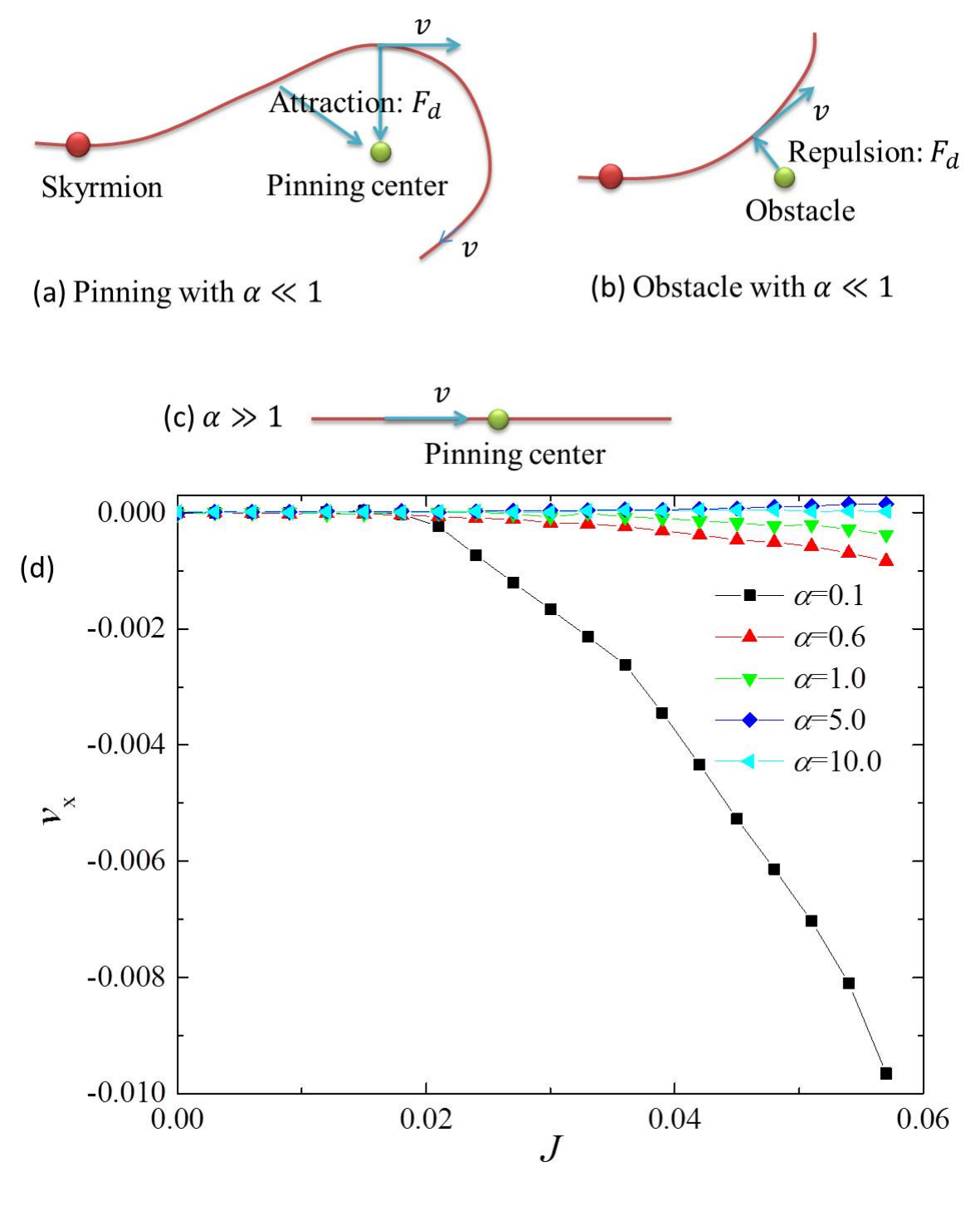,width=\columnwidth}
\caption{\label{f7}(color online) (a) and (b): Schematic view of a skyrmion passing through a pinning center (a) and obstacle (b) when the Magnus force is dominant. When the Magnus force is dominant over the dissipative force, the skyrmion is deflected by the pinning centers or obstacles. (c) Same as (a) and (b) except that the dissipative force is dominant. The skyrmion has to overcome the pinning site or obstacle by passing through it. (d) Numerical results of current-velocity curves with different $\alpha$. The depinning current increases with $\alpha$.}
\end{figure}

It is also interesting to discuss the effect of the Magnus force on the pinning of skyrmions.  The Magnus force dominates over the dissipative force, $F_M\gg 4\pi\alpha v/ \gamma$, for $\alpha\ll 1$. When a skyrmion moves around a pinning center or an obstacle, it is easily scattered with a velocity perpendicular to the pinning force or repulsive force. Thus, the skyrmion avoids passing through the pinning center and its influence  is minimized, as shown in Fig.~\ref{f7} (a) and (b). When the dissipative force is dominant, $\alpha \gg 1$, the skyrmion has to pass through the pinning center so the pinning becomes very strong, as it shown in Fig.~\ref{f7}(c). We performed numerical simulations for several  $\alpha$ ratios [see Fig. \ref{f7} (d)]. The depinning current is weaker for smaller values of $\alpha$, i.e., when the Magnus force is dominant.

\section{Discussion}
The equation of motion for the center of mass of a rigid skyrmion lattice has been derived in Ref.~\onlinecite{Everschor12} by using an approach proposed by Thiele~\cite{Thiele72}. This equation is similar to Eq.~\eqref{eq5} for $F_{ss}=0$ and $F_{d}=0$. The equation of motion for the collective excitations in the skyrmion lattice was derived in Ref.~\onlinecite{Zang11}, and it also shares a similar structure with Eq.~\eqref{eq5}. The equation of motion presented in this work is more general because it also describes the dynamics of a single skyrmion and its interaction with other skyrmions and defects.

The resulting equation of motion is similar to that of vortices in type II superconductors. However, the pinning of superconducting vortices is much stronger than that of skyrmions because of the following reasons. The Magnus force is negligibly small for vortices, except in the super clean region, which has not yet been realized experimentally. \cite{Blatter94}  In contrast, the Magnus force of skyrmions is stronger than the dissipative forces. Vortices have to pass through the pinning centers, which leads to large critical currents, while skyrmions can be easily deflected by the pinning center because of the dominant Magnus force. \cite{Iwasaki2013} In addition, defects suppress superconductivity over a range that is equal or larger than the superconducting coherence length (linear size of the normal vortex core). This matching of length scales makes the pinning rather strong. In contrast, the characteristic length of defects (inter-atomic spacing) is much smaller than the core size of the skyrmions.

\section{Conclusion}
We have derived an effective particle model for skyrmions, which includes repulsive skyrmion-skyrmion interactions, interaction with defects and the role of the Magnus force. The model successfully describes the rotation of the skyrmion lattice in the presence of a temperature gradient and explains the small depinning thresholds that have been  experimentally observed. It also provides clear predictions for quantum and thermal creep. Finally the model has been validated by direct comparisons of the results of depinning and transport curves against the original continuum model. Our particle model offers a convenient and transparent theoretical framework for the future computational and analytical studies of skyrmions. 

\section{Acknowledgments}
We thank Christian Pfleiderer, Shinichiro Seki, A. N. Bogdanov, Ivar Martin, Yasuyuki Kato, Leonardo Civale and Boris Maiorov for useful discussions and Cynthia Reichhardt for a critical reading of the manuscript. This work was supported by the US Department of Energy, Office of Basic Energy Sciences, Division of Materials Sciences and Engineering, and was carried out under the auspices of the NNSA of the US DoE at LANL under Contract No. DE-AC52-06NA25396.

\appendix

\section{Numerical details}
In simulations, we introduce dimensionless units in Eqs. \eqref{eq3} and \eqref{eq4}. Length is in unit of $J_{\rm{ex}}/D$; energy is in unit of $J_{\rm{ex}}^2/D$; magnetic field is in unit of $D^2/J_{\rm{ex}}$; time is in unit of $J_{\rm{ex}}/(\gamma D^2)$; current is in unit of $2D e/\hbar$. We use the periodic boundary condition in both directions. To find the ground state, we anneal the system by adding a Gaussian noisy magnetic field along the $z$ direction in $\mathbf{H}_{\rm{eff}}$. Equation \eqref{eq4} is solved by an explicit numerical scheme developed in Ref. \onlinecite{Serpico01}. The current is along the $x$ direction. In calculations of the results in Fig. \ref{f3}, we use a simulation box of size $L_x\times L_y=30\times 10$. The system is discretized with a grid size of $0.2$. In calculations of the results in Fig. \ref{f6}, the defects are modeled by Eq. \eqref{eq11} with $J_d=1.0$ and $\xi_d=1.0$, where the interaction between skyrmions and defects is repulsive. The $N_d=500$ defects are randomly distributed in a simulation box of size $L_x\times L_y=100\times 100$. The I-V curves are obtained by averaging over $20$ realizations of random defects.

In the particle-level simulation, we take the interaction between skyrmions as $\mathbf{F}_{ss}=F_{s0} K_1(r_d/\xi)\hat{r}_d$ and the repulsive interaction between skyrmions and defects as $\mathbf{F}_d=F_{d0}\exp(-r_d/\xi_d)\hat{r}_d$. Here $\hat{r}_d$ is a unit vector along $r_d$. In dimensionless units, the equation of motion becomes
\begin{equation}\label{eqm1}
\alpha \mathbf{v}_i= \sum_j^N \mathbf{F}_{ss}(\mathbf{r}_i-\mathbf{r}_j)+\sum_j^{N_d}\mathbf{F}_d (\mathbf{r}_i-\mathbf{r}_{d,j})+\hat{z}\times \mathbf{J}+\hat{z}\times\mathbf{v}_i.
\end{equation}
In simulation $\xi_d=2\xi=2.0$, $F_{s0}=1.0$ and $F_{d0}=0.6$. The number of skyrmions $N$ and the number of defects $N_d$ are $N=N_d=225$. The simulation box is $L_x\times L_y=45\times 39$. We use the second-order Runge-Kutta method to integrate Eq. \eqref{eqm1} with a time step $\Delta t=0.05$.

\section{Thermal activation of a skyrmion over a barrier}
Here we calculate the thermal activation rate for a skyrmion in a metastable potential $U(x,y)$. The equation of motion for the skyrmion in dimensionless units is
\begin{equation}\label{eqs1}
\tilde{\mathbf{f}}+\left[\hat{z}\times \mathbf{v}-\nabla U\right]=\alpha  \mathbf{v},
\end{equation}
with a Gaussian noisy force $\left\langle \tilde{f}_\mu(\mathbf{r},t)\right\rangle=0$ and 
\begin{equation}\label{eqs2}
\left\langle \tilde{f}_\mu(\mathbf{r},t) \tilde{f}_{\mu'}(\mathbf{r}',t')\right\rangle =2 \alpha  T \delta  \left(\mathbf{r}-\mathbf{r}'\right) \delta  \left(t-t'\right)\delta_{\mu, \mu'}.
\end{equation}
Using the nonequilibrium path integral approach \cite{SimonsQFT}, the probability of finding the skyrmion at $\mathbf{r}'$ at $t'$ starting from the initial position $\mathbf{r}_0$ at $t_0$ is
\begin{equation}\label{eqs3}
p(\mathbf{r'}, t'|\mathbf{r}_0,t_0) = \int\mathcal{D}[\mathbf{r},\tilde{\mathbf{r}}]\exp \left(\int dt\mathcal {L}\right),
\end{equation}
\begin{equation}\label{eqs4}
{\cal L} = i{\tilde{\mathbf{r}}}\cdot\left[{\partial _t}{\bf{r}} - \left( - \frac{1}{\alpha }{\partial _r}U + \frac{1}{\alpha }{\hat{z}\times\mathbf{v}}\right)\right] - \frac{\tilde{T}}{2}{{\tilde{\mathbf{r}}}^2},
\end{equation}
with $\tilde{T}={2 T}/{\alpha }$. It is more convenient to use the Hamiltonian description
\begin{align}
\nonumber\mathcal{H}=\mathbf{p}\cdot \partial_t\mathbf{r}-\mathcal{L}=\\
\nonumber-\left[\frac{A \alpha ^2 \left(p_x^2+p_y^2\right)}{2 \left(\alpha ^2+1\right)}+\frac{{\partial_x U} \left(\alpha  p_x+p_y\right)}{\alpha ^2+1}+\frac{{\partial_y U} \left(\alpha  p_y-p_x\right)}{\alpha ^2+1}\right]
\end{align}
where the conjugate momentum $\mathbf{p}$ is defined as $p_x=\partial \mathcal{L}/\partial v_x$ and $p_y=\partial \mathcal{L}/\partial v_y$. For a weak noise $\tilde{T}\ll 1$, the dominant contribution to the path integral are those trajectories governed by the standard Hamiltonian dynamics. For the Hamiltonian dynamics, $\mathcal{H}$ is conserved. Initially for the skyrmion at the well $\mathbf{r}_w$, the system has $\mathcal{H}=0$ since $p_x=p_y=0$. We then look for the trajectories with $\mathcal{H}=0$ and with minimal action $S_T=-\int dt \mathcal{L}$. One obvious solution for $\mathcal{H}=0$ is $p_x=p_y=0$ with $S_T=0$. This is a non-fluctuating trajectory, which does not contribute to the thermal activation of skyrmions. There is another trajectory with a minimal $S_T$
\begin{equation}\label{eqs5}
p_\mu=-\frac{2}{\tilde{T}\alpha}\partial_\mu U,
\end{equation}
with $\mu=x,\ y$. The corresponding equation of motion is
\begin{equation}\label{eqs6}
\left[\hat{z}\times \mathbf{v}-\nabla U\right]=-\alpha  \mathbf{v}.
\end{equation}
Compared with Eq. \eqref{eqs1}, the sign of damping is changed. Thus Eq. \eqref{eqs6} describes the motion of a skyrmion in a potential with a negative damping, which forces the skyrmion to leave the well, and contributes to the thermal activation. The action $S_T$ for this trajectory is 
\begin{equation}\label{eqs7}
S_T=\int dt \mathbf{p}\cdot \mathbf{v}=\frac{2}{\alpha \tilde{T}}\int_{\mathbf{r}_w}^{\mathbf{r}_b} [ \nabla U\cdot d\mathbf{r}]=\frac{\Delta U}{T},
\end{equation} 
where the integration is from the well $\mathbf{r}_w$ to the barrier $\mathbf{r}_b$ of the potential $U$, and $\Delta U=U(\mathbf{r}_b)-U(\mathbf{r}_w)$ is the height of the barrier. The action does not depend on the Magnus force. The reasons are as follows: first the dynamics of skyrmions is irrelevant for the thermal activation of skyrmions over the barrier. The probability distribution of skyrmion in the potential only depends on the potential energy. Secondly, the Magnus force does not produce work when skyrmions move.

%

\end{document}